\documentclass[10pt]{article}
\usepackage{graphicx}
\usepackage{amsmath}
\usepackage{amssymb}
\usepackage{caption2}
\begin{document}
\bibliographystyle{prsty}
\begin{center}
{\large {\bf \sc{Two-pion transitions from the $P$-wave to $S$-wave charmonium states  }}} \\[2mm]
Zhi-Gang Wang  \footnote{E-mail:wangzgyiti@yahoo.com.cn. } \\
  Department of Physics, North China Electric Power University, Baoding 071003, P. R.
  China
\end{center}

\begin{abstract}
In this article, we study the two-pion decays  in the $P$-wave to $S$-wave charmonium
transitions with the heavy meson  effective theory, and make qualitative predictions
for the ratios among the two-pion decay widths.
\end{abstract}


\section{Introduction}

Recently, the  BESIII collaboration searched for the two-pion transitions  $\chi_{cj}\to \eta_{c}\pi^{+}\pi^{-}$ ($j=0, 1, 2$)  using a sample of $1.06\times 10^{8}$ $\psi(3686)$ events collected by the BESIII detector, and observed no signals for  the three $\chi_{cj}$ states in the  $\eta_c$ decay modes, and set   the upper limits for the branching ratios ${\rm Br}(\chi_{c0}\to \eta_{c}\pi^{+}\pi^{-})<0.07\%$, ${\rm Br}(\chi_{c1}\to \eta_{c}\pi^{+}\pi^{-})<0.32\%$,   ${\rm Br}(\chi_{c2}\to \eta_{c}\pi^{+}\pi^{-})<0.54\%$ at the $90\%$ confidence level \cite{BES1208}. Taking into account the widths of the $P$-wave charmonium states $\Gamma(\chi_{c2})=(1.95\pm0.13)\,\rm{MeV}$, $\Gamma(\chi_{c1})=(0.88\pm0.05)\,\rm{MeV}$, $\Gamma(\chi_{c0})=(10.5\pm0.8)\,\rm{MeV}$ from the Particle Data Group \cite{PDG}, we can obtain the two-pion  decay widths $\Gamma(\chi_{c2}\to \eta_{c}\pi^{+}\pi^{-})<10.53\,\rm{KeV}$, $\Gamma(\chi_{c1}\to \eta_{c}\pi^{+}\pi^{-})<2.186\,\rm{KeV}$ and $\Gamma(\chi_{c0}\to \eta_{c}\pi^{+}\pi^{-})<7.35\,\rm{KeV}$.
While in previous studies, the BABAR collaboration searched for the processes $\gamma \gamma \to X \to \eta_c \pi^+ \pi^-$, where the $X$ stands for  the resonances $\chi_{c2}$, $\eta_c^{\prime}$, $\chi_{c2}^{\prime}$, $X(3872)$, $X(3915)$,  and set the   upper limit ${\rm Br}(\chi_{c2}\to \eta_{c}\pi^{+}\pi^{-})<2.2\%$  at
the $90\%$ confidence level  \cite{BABAR1206}. For the most promising process   $\chi_{c1}\to \eta_c\pi^+\pi^-$  dominated
by the $E_1-M_1$ transition, the upper limit  ${\rm Br}(\chi_{c1}\to \eta_{c}\pi^{+}\pi^{-})<0.32\%$ is lower than the existing theoretical prediction ${\rm Br}(\chi_{c1}\to \eta_c\pi\pi) =
(2.72\pm 0.39)\%$ based on the QCD multipole expansion by almost an order of magnitude \cite{BES1208,Kuang2007}. In the QCD multipole expansion,
the $\chi_{c1}\to \eta_c\pi^+\pi^-$ branching ratio (or width) dominated by the $E_1-M_1$ transition is significantly smaller than that of the $E_1-E_1$ transitions, for example, the transition $\psi^{\prime} \to J/\psi \pi^+ \pi^-$ (or $\eta_c^{\prime} \to \eta_c \pi^+\pi^-$) \cite{PDG,WangEPJA}.
In Ref.\cite{Voloshin1209}, Voloshin studies  the charmonium transitions  $\chi_{cj}\to\eta_c$ ($j=0,1,2$) with emission of one or two pions, and shows that only   the  decays $\chi_{c1} \to \eta_c \pi^+ \pi^-$ and $\chi_{c0} \to \eta_c \pi^0$ take place  in the leading order in the QCD multipole expansion, and obtains the predictions  $\frac{\Gamma(\chi_{c0}\to \eta_c \pi^0)}{\Gamma( \chi_{c1}\to\eta_c\pi^+\pi^-)}\approx 13.7$ and $\frac{\Gamma(h_{c}\to J/\psi \pi^+\pi^-)}{\Gamma (\chi_{c1}\to\eta_c\pi^+\pi^-)}\approx 0.1$.
In this article, we study the two-pion decays in the $P$-wave to $S$-wave charmonium
transitions   with  the heavy meson effective theory in the leading
order approximation. The heavy meson effective  theory has been successfully applied to
  study the pseudoscalar meson decays of the charmed mesons \cite{HMET-id}, and the radiative and vector-meson decays of the heavy quarkonium states \cite{HMET-RV}.

The article is arranged as follows:  we study the two-pion decays in the $P$-wave to $S$-wave charmonium
transitions  with the heavy meson effective
theory in Sect.2; in Sect.3, we present the
 numerical results and discussions; and Sect.4 is reserved for our
conclusions.

\section{ The two-pion transitions  with the heavy meson effective theory }

  The charmonium states have the same radial
quantum number $n$ and  orbital angular  momentum $L=0,1$ can be expressed
by  the superfields $J$, $J^\mu$ \cite{GattoPLB93,PRT1997},
\begin{eqnarray}
J&=&\frac{1+{\rlap{v}/}}{2}\left\{\psi_{\mu}\gamma^\mu-\eta_c\gamma_5\right\}
\frac{1-{\rlap{v}/}}{2} \, , \nonumber \\
J^\mu&=&\frac{1+{\rlap{v}/}}{2}\left\{\chi_{c2}^{\mu\nu}\gamma_\nu+\frac{1}{\sqrt{2}}\epsilon^{\mu\alpha\beta\lambda}v_\alpha
\gamma_{\beta}\chi^{c1}_{\lambda}+\frac{1}{\sqrt{3}}\left(\gamma^\mu-v^\mu\right)\chi_{c0}+h^\mu_c\gamma_5\right\}
\frac{1-{\rlap{v}/}}{2} \, ,
\end{eqnarray}
where the $v^{\mu}$ denotes the four-velocity associated to the
superfields, the charmonium states $\chi_{c2}^{\mu\nu}$,
$\chi_{c1}^{\mu}$, $\chi_{c0}$, $h_{c}^{\mu}$, $\psi^\mu$, $\eta_c$ have the total
angular momentum $j=2,1,0,1,1,0$,  respectively, and belong to different
multiplets.  We multiply  those charmonium fields
  with  the factors $\sqrt{M_{\chi_{c2}}}$, $\sqrt{M_{\chi_{c1}}}$, $\sqrt{M_{\chi_{c0}}}$, $\sqrt{M_{h_c}}$, $\sqrt{M_{\psi}}$, $\sqrt{M_{\eta_{c}}}$, respectively,  and they have dimension of mass $\frac{3}{2}$.  The superfields
$J^{(\mu)}$   have the following properties under the
parity ($P$), charge conjunction ($C$), heavy quark spin transformations ($S$),
\begin{eqnarray}
J^{(\mu)}&\stackrel{P}{\longrightarrow}&\gamma^{0}J_{(\mu)}\gamma^{0} \, ,  \nonumber \\
J^{(\mu)}&\stackrel{C}{\longrightarrow}&(-1)^{L+1}C[J_{(\mu)}]^{T}C \,,\nonumber \\
J^{(\mu)}&\stackrel{S}{\longrightarrow}&S J_{(\mu)}S^{\prime\dagger}\,,\nonumber \\
v^{\mu}&\stackrel{P}{\longrightarrow}&v_{\mu}\, ,
\end{eqnarray}
where $S,S^{\prime}\in SU(2)$ heavy quark spin symmetry groups, and
$[S,{\rlap{v}/}]=[S^{\prime},{\rlap{v}/}]=0$.

The $\pi^+\pi^-$ transitions between the $P$-wave and $S$-wave
charmonium states  can be described by the following
phenomenological Lagrangians ${\cal{L}}_{1}$ and ${\cal{L}}_{2}$,
\begin{eqnarray}
{\cal{L}}_{1}&=&\frac{g_c}{\Lambda}\mathrm{Tr}\left[J_\mu \sigma^{\mu\rho}\bar{J}\right]v^\beta \mathrm{Tr} \left[{\cal{A}_\rho} {\cal{A}_\beta}\right]
-\frac{ig_d}{\Lambda} \mathrm{Tr}\left[J_\mu \sigma^{\mu\tau}\bar{J}\sigma_{\tau\rho}\right]v^\beta  \mathrm{Tr} \left[{\cal{A}^\rho} {\cal{A}_\beta}\right] +h.c.\, , \\
{\cal{L}}_{2}&=&-\frac{ig_f}{\Lambda^2}\mathrm{Tr}\left[J_\mu \sigma^{\lambda\beta}\partial_{\lambda}\bar{J}\right]  \mathrm{Tr} \left[{\cal{A}^\mu} {\cal{A}_\beta}\right]
-\frac{g_h}{\Lambda^2} \mathrm{Tr}\left[J_\mu \sigma^{\lambda\tau}\partial_{\lambda}\bar{J}\sigma_{\tau\beta}\right]  \mathrm{Tr} \left[{\cal{A}^\mu} {\cal{A}^\beta}\right] +h.c. \, ,
\end{eqnarray}
where
\begin{eqnarray}
{\cal{A}_\mu}&=&\frac{1}{2}\left(\xi^{\dag}\partial_\mu \xi-\xi\partial_\mu\xi^{\dag}\right)=\frac{i\partial_\mu{\cal M}}{f_\pi}+\cdots \, ,
\end{eqnarray}
and $\bar{J}=\gamma^0 J^{\dag} \gamma^0$. We introduce an energy scale $\Lambda=1\,\rm{GeV}$ to warrant that the strong coupling constants $g_c$, $g_d$, $g_f$ and $g_h$ are dimensionless quantities.
The light pseudoscalar mesons are described by the fields
 $\displaystyle \xi=\exp\left(i {\cal M} \over f_\pi \right)$, where
\begin{equation}
{\cal M}= \left(\begin{array}{ccc}
\sqrt{\frac{1}{2}}\pi^0+\sqrt{\frac{1}{6}}\eta & \pi^+ & K^+\\
\pi^- & -\sqrt{\frac{1}{2}}\pi^0+\sqrt{\frac{1}{6}}\eta & K^0\\
K^- & {\bar K}^0 &-\sqrt{\frac{2}{3}}\eta
\end{array}\right) \, ,
\end{equation}
and the decay constant $f_\pi=130\,\rm{MeV}$.
  The Lagrangians ${\cal{L}}_{1}$ and ${\cal{L}}_{2}$   violate the heavy quark spin symmetry, and describe the $E_1-M_1$ and $E_1-M_2$ transitions between the $P$-wave and $S$-wave charmonium states, respectively. For review of the QCD multipole expansion, one can consult Ref.\cite{ReviewVoloshin}.  The Lagrangian ${\cal{L}}_{1}$ is taken from Ref.\cite{GattoPLB93} and  the Lagrangian  ${\cal{L}}_{2}$ is constructed in this article.
We carry out the trace in the heavy meson effective Lagrangians  ${\cal L}_{1}$  and ${\cal L}_{2}$, and observe that the  decays $\chi_{c2}\to J/\psi \pi^+\pi^-$, $\chi_{c1}\to J/\psi \pi^+\pi^-$, $\chi_{c0}\to J/\psi \pi^+\pi^-$, $h_{c}\to J/\psi \pi^+\pi^-$, $\chi_{c1}\to \eta_c \pi^+\pi^-$, $h_{c}\to \eta_c \pi^+\pi^-$ receive contributions from both the    $E_1-M_1$ and $E_1-M_2$ transitions, while the decay $\chi_{c2}\to \eta_c \pi^+\pi^-$ only receives contribution from the   $E_1-M_2$ transition.
The charmonium parts $\mathrm{Tr}\left[J_\mu \sigma^{\mu\rho}\bar{J}\right]$, $\mathrm{Tr}\left[J_\mu \sigma^{\mu\tau}\bar{J}\sigma_{\tau\rho}\right]$,
$\mathrm{Tr}\left[J_\mu \sigma^{\lambda\beta}\partial_{\lambda}\bar{J}\right]$ and $\mathrm{Tr}\left[J_\mu \sigma^{\lambda\tau}\partial_{\lambda}\bar{J}\sigma_{\tau\beta}\right]$ in the effective  Lagrangians lead to the spin-conserved (electro-like) transitions $\chi_{cj}\to J/\psi$, $h_c \to \eta_c$ and spin-violated (magnetic-like) transitions $\chi_{c1} \to \eta_c$, $h_c \to J/\psi$. The electro-like ($E_1$) and magnetic-like ($M_1$, $M_2$) processes manifest themselves through $\chi_{cj}\to J/\psi$, $h_c \to \eta_c$ and $\chi_{c1} \to \eta_c$, $h_c \to J/\psi$ respectively in the  $E_1-M_1$ (or  $E_1-M_2$) transitions, just like the children always manifest the feature of one parent in biology; the spin-conserved transitions violate isospin symmetry while the spin-violated transitions conserve isospin symmetry. The decays $\chi_{cj}\to J/\psi \pi^+\pi^-$ and $h_c \to \eta_c \pi^+\pi^-$ violate  isospin symmetry or $G$-parity, and can also take place through the $E_1$ electromagnetic interactions with the intermediate $\rho$ meson, this mechanism corresponds to the  effective Lagrangian  $g_{e}\mathrm{Tr}\left[J^{\alpha}\bar{J}\right]v^\beta \mathrm{Tr} \left[{\cal{A}_\alpha} {\cal{A}_\beta}\right]$, the charmonium part $\mathrm{Tr}\left[J^{\alpha}\bar{J}\right]$ only leads to the spin-conserved transitions $\chi_{cj}\to J/\psi$ and $h_c \to \eta_c$.   We can draw the conclusion tentatively that the electro and electro-like interactions lead to the isospin-violated decays, and the $g_{e}\mathrm{Tr}\left[J^{\alpha}\bar{J}\right]v^\beta \mathrm{Tr} \left[{\cal{A}_\alpha} {\cal{A}_\beta}\right]$ can be  absorbed  in the effective  Lagrangian  ${\cal L}_{1}$. The isospin-violated processes can take place, for example, the BESIII collaboration had precisely measured the branching ratio  ${\rm{Br}}(\psi^{\prime}\to J/\psi\pi^0)=(1.26\pm0.02\pm0.03)\times 10^{-3}$ \cite{BES-psi-pi}.

We write down the   two-pion transition amplitudes $T$, and obtain the decay widths $\Gamma$,
\begin{eqnarray}
\Gamma&=&\frac{1}{(2j+1)2M_{i}}\int \sum|T|^2\frac{dl^2}{2\pi}d\Phi(P\to q,l)d\Phi(l\to r,t) \, ,
\end{eqnarray}
where the $d\Phi(P\to q,l)$ and $d\Phi(l\to r,t)$ are the two-body phase factors,
\begin{eqnarray}
d\Phi(P\to q,l)&=&(2\pi)^4\delta^4(P-q-l)\frac{d^3\vec{l}}{(2\pi)^3 2l_0}\frac{d^3\vec{q}}{(2\pi)^3 2q_0} \, , \nonumber\\
d\Phi(l\to r,t)&=&(2\pi)^4\delta^4(l-r-t)\frac{d^3\vec{r}}{(2\pi)^3 2r_0}\frac{d^3\vec{t}}{(2\pi)^3 2t_0} \, ,
\end{eqnarray}
the $P_\mu$ and $q_\mu$ are the momenta of the initial and final charmonium states respectively, the $M_i$ is the mass of the initial charmonium state, the $j$
is the total spin of the initial charmonium state, the $r_\mu$ and $t_\mu$
are the momenta of the $\pi^+$ and $\pi^-$ respectively, and the $l^2$ is the invariant mass of the $\pi^+\pi^-$ system. We carry out
the integrals $d\Phi(l\to r,t)$, $d\Phi(P\to q,l)$ and $d l^2$ sequentially. In calculations, we have used the following formula,
\begin{eqnarray}
\int t_{\alpha} t_{\mu} r_{\beta} r_{\nu} d\Phi(l\to r,t) &=& \frac{\vec{t}}{4\pi \sqrt{l^2}} \left\{A(g_{\alpha\mu}g_{\beta\nu}+g_{\alpha\beta}g_{\mu\nu}+g_{\alpha\nu}g_{\beta\mu}) +B\frac{g_{\alpha\mu}l_{\beta}l_{\nu}+g_{\beta\nu}l_{\alpha}l_{\mu}}{l^2}\right.\nonumber\\
&&\left.+C\frac{g_{\alpha\beta}l_{\mu}l_{\nu}+g_{\mu\nu}l_{\alpha}l_{\beta}+g_{\alpha\nu}l_{\beta}l_{\mu}+g_{\beta\mu}l_{\alpha}l_{\nu}}{l^2}+D\frac{l_{\alpha}l_{\mu}l_{\beta}l_{\nu}}{l^4} \right\} \, ,
\end{eqnarray}
where
\begin{eqnarray}
A&=&\frac{l^4-8l^2m_\pi^2+16m_\pi^4}{240} \, , \nonumber \\
B&=&-\frac{3l^4-14l^2m_\pi^2+8m_\pi^4}{120} \, , \nonumber \\
C&=&\frac{l^4-3l^2m_\pi^2-4m_\pi^4}{60} \, , \nonumber \\
D&=&\frac{l^4+2l^2m_\pi^2+6m_\pi^4}{30} \, ,
\end{eqnarray}
and we  carry out the sum of all the polarization  vectors of the charmonium states  using the FeynCalc.

\section{Numerical Results}
We take the input parameters  from the Particle Data Group \cite{PDG},
$M_{\chi_{c2}}=3556.20\,\rm{MeV}$, $M_{\chi_{c1}}=3510.66\,\rm{MeV}$,  $M_{\chi_{c0}}=3414.75\,\rm{MeV}$, $M_{h_{c}}=3525.41\,\rm{MeV}$, $M_{J/\psi}=3096.916\,\rm{MeV}$, $M_{\eta_{c}}=2981.0\,\rm{MeV}$, and $M_{\pi}=139.57\,\rm{MeV}$, and obtain
the numerical values of the two-pion decay widths,
\begin{eqnarray}
\Gamma(\chi_{c2}\to J/\psi \pi^+\pi^-)&=&\left\{43.8353(g_c+g_d)^2+3.7406(g_c+g_d)g_f +9.6279(g_c+g_d)g_h\right.\nonumber\\
       &&\left.+0.3957g_f^2+0.5224g_{f} g_{h}+2.9811g_h^2 \right\} \,\rm{KeV}\, ,\nonumber \\
\Gamma(\chi_{c1}\to J/\psi \pi^+\pi^-)&=&\left\{14.2933(g_c+g_d)^2+0.9305(g_c+g_d)(g_f-g_h)+0.0546(g_f-g_h)^2 \right\} \,\rm{KeV}\, ,\nonumber\\
\Gamma(\chi_{c0}\to J/\psi \pi^+\pi^-)&=&\left\{0.7121g_c^2-2.8483g_{c}g_d -0.0136g_{c}g_f-0.0347g_{c}g_h+2.8483g_d^2-0.0273g_{d}g_f\right.\nonumber\\
       &&\left.+0.0693g_{d}g_h+0.0001g_f^2-0.0002g_{f} g_{h}+0.0009g_h^2 \right\} \,\rm{KeV}\, ,\nonumber \\
\Gamma(h_{c}\to J/\psi \pi^+\pi^-)&=&\left\{42.6521(g_c-g_d)^2+3.0694(g_c-g_d)(g_f-g_h)+0.1978(g_f-g_h)^2 \right\} \,\rm{KeV}\, ,\nonumber\\
\Gamma(\chi_{c2}\to \eta_c \pi^+\pi^-)&=&\left\{3.3352(g_f+g_h)^2 \right\} \,\rm{KeV}\, ,\nonumber \\
\Gamma(\chi_{c1}\to \eta_c \pi^+\pi^-)&=&\left\{340.7208(g_c+g_d)^2+40.3081(g_c+g_d)(g_f+g_h)+2.0529(g_f+g_h)^2 \right\} \,\rm{KeV}\, ,\nonumber\\
\Gamma(h_{c}\to \eta_c \pi^+\pi^-)&=&\left\{868.4603g_d^2+356.3769g_{d} g_{h}+52.2813g_h^2 \right\} \,\rm{KeV}\, .
\end{eqnarray}
In general, we expect to fit  the coupling constants $g_c$, $g_d$, $g_f$ and $g_h$  to the precise experimental data, however, in the
 present time the experimental data are rare.

In the QCD multipole expansion, the Hamiltonians for the chromo-electric dipole $E_1$, the
chromo-magnetic dipole $M_1$ and the chromo-magnetic quadrupole $M_2$ transitions  are
\begin{eqnarray}
H_{E_1}&=&-\frac{1}{2}\xi^a \vec{r}\cdot \vec{E}^a \, , \nonumber\\
H_{M_1}&=&-\frac{1}{2m_Q}\xi^a \vec{\Delta}\cdot \vec{B}^a \, , \nonumber\\
H_{M_2}&=&-\frac{1}{4m_Q}\xi^a S_jr_i (D_i B_j(0))^a \, ,
\end{eqnarray}
where the $\xi^a=t^a_1-t^a_2$ is the difference of the color generators acting on the quark and antiquark,
the $\vec{\Delta}=\vec{\sigma}_1-\vec{\sigma}_2$ is the spin operator  with
$\vec{\sigma}_1$ and $\vec{\sigma}_2$ acting on the quark and  antiquark respectively,
the $\vec{r}$ is the relative vector   of the quark and
 antiquark,  the $\vec{S}=\frac{\vec{\sigma}_1+\vec{\sigma}_2}{2}$   is the operator of the total spin of the quark-antiquark pair, and the $\vec{D}$ is the
QCD covariant derivative, the $\vec{E}^a$ and $\vec{B}^a$ are the chromo-electric and chromo-magnetic components
of the gluon field strength tensor respectively \cite{ReviewVoloshin}. The $E_1-M_1$ and $E_1-M_2$ transitions both take place at the next-to-leading  order ${\cal O}(\frac{1}{m_Q})$,  they are suppressed compared to  the $E_1-E_1$ transitions, which take place in the leading order ${\cal O}(1)$. There is an additional  covariant derivative $\vec{D}$ in the $H_{M_2}$, the $E_1-M_2$ transitions are suppressed in the phase-space compared to  the $E_1-M_1$ transitions, if the momentum transfer in the $P$-wave to $S$-wave charmonium transitions  is small. From Eq.(11), we can see that the coefficients (which can be denoted as $X$) of the coupling constants $g_c$, $g_d$, $g_f$ and $g_h$ have the  hierarchy $X(g_c^2),|X(g_{c}g_d)|,X(g_d^2)\gg |X(g_{c}g_f)|, |X(g_{d}g_f)|, |X(g_{c}g_h)|, |X(g_{d}g_h)|\gg X(g_f^2), |X(g_{f}g_h)|, X(g_h^2)$, the $E_1-M_2$ transitions are suppressed indeed in the phase-space.

If we take the approximation $g_f\approx g_h \approx 0$, i.e. neglect the $E_1-M_2$ transitions, the two-pion decay widths can be simplified as
\begin{eqnarray}
\Gamma(\chi_{c2}\to J/\psi \pi^+\pi^-)&=& 43.8353(g_c+g_d)^2  \,\rm{KeV}\, ,\nonumber \\
\Gamma(\chi_{c1}\to J/\psi \pi^+\pi^-)&=& 14.2933(g_c+g_d)^2  \,\rm{KeV}\, ,\nonumber\\
\Gamma(\chi_{c0}\to J/\psi \pi^+\pi^-)&=& 0.7121g_c^2-2.8483g_{d}(g_c-g_d) \,\rm{KeV}\, ,\nonumber \\
\Gamma(h_{c}\to J/\psi \pi^+\pi^-)&=& 42.6521(g_c-g_d)^2  \,\rm{KeV}\, ,\nonumber\\
\Gamma(\chi_{c2}\to \eta_c \pi^+\pi^-)&=&0 \,\rm{KeV}\, ,\nonumber \\
\Gamma(\chi_{c1}\to \eta_c \pi^+\pi^-)&=& 340.7208(g_c+g_d)^2  \,\rm{KeV}\, ,\nonumber\\
\Gamma(h_{c}\to \eta_c \pi^+\pi^-)&=& 868.4603g_d^2 \,\rm{KeV}\, ,
\end{eqnarray}
then we obtain the ratios
\begin{eqnarray}
\frac{\Gamma(\chi_{c2}\to J/\psi \pi^+\pi^-)}{\Gamma(\chi_{c1}\to \eta_c \pi^+\pi^-)}&=&0.129 \, , \, 0.129\, , \, 0.129\, , \nonumber\\
\frac{\Gamma(\chi_{c1}\to J/\psi \pi^+\pi^-)}{\Gamma(\chi_{c1}\to \eta_c \pi^+\pi^-)}&=&0.042 \, , \,0.042\, ,\, 0.042\, ,  \nonumber\\
\frac{\Gamma(\chi_{c0}\to J/\psi \pi^+\pi^-)}{\Gamma(\chi_{c1}\to \eta_c \pi^+\pi^-)}&=&0.001  \, , \, 0.002 \, ,\, 0.008\, , \nonumber\\
\frac{\Gamma(h_{c}\to J/\psi \pi^+\pi^-)}{\Gamma(\chi_{c1}\to \eta_c \pi^+\pi^-)}&=&0.0 \, , \, 0.125 \, , \, 0.125 \, , \nonumber\\
\frac{\Gamma(h_{c}\to \eta_c \pi^+\pi^-)}{\Gamma(\chi_{c1}\to \eta_c \pi^+\pi^-)}&=&0.637\, , \, 0.0\, , \, 2.549  \, ,
\end{eqnarray}
with  the additional approximation $g_c=g_d$, $g_d=0$, $g_c=0$, respectively. The present prediction   $\frac{\Gamma(h_{c}\to J/\psi \pi^+\pi^-)}{\Gamma(\chi_{c1}\to \eta_c \pi^+\pi^-)}=0.125$ with the coupling $g_c=0$ (or $g_d=0$) is consistent with the value $0.1$ from the QCD multipole expansion \cite{Voloshin1209}. By measuring the ratio  $\frac{\Gamma(h_{c}\to \eta_c \pi^+\pi^-)}{\Gamma(\chi_{c1}\to \eta_c \pi^+\pi^-)}$, we can obtain powerful constraint on the couplings $g_c$ and $g_d$.
 If the value $g_c=0$ is excluded, the transition $\chi_{c1}\to \eta_c \pi^+\pi^-$ is the most promising process, and the transitions $\chi_{c0}\to J/\psi \pi^+\pi^-$, $h_{c}\to J/\psi \pi^+\pi^-$, $\chi_{c2}\to \eta_c \pi^+\pi^-$ are greatly suppressed.
The upper limits of the two-pion  decay widths $\Gamma(\chi_{c2}\to \eta_{c}\pi^{+}\pi^{-})<10.53\,\rm{KeV}$ and $\Gamma(\chi_{c1}\to \eta_{c}\pi^{+}\pi^{-})<2.186\,\rm{KeV}$ lead to the prediction $|g_f+g_h|<1.7769$ and $|g_c+g_d|<0.0801$ from the formulae
$\Gamma(\chi_{c2}\to \eta_c \pi^+\pi^-)= 3.3352(g_f+g_h)^2   \,\rm{KeV}$
and $\Gamma(\chi_{c1}\to \eta_c \pi^+\pi^-)= 340.7208(g_c+g_d)^2   \,\rm{KeV}$, respectively, which are inconsistent with our naive expectation based on the QCD multipole expansion. Precise measurements are needed. The   transitions $h_{c}\to J/\psi \pi^+\pi^-$ and $\chi_{c1}\to \eta_c \pi^+\pi^-$ are of particular interesting, the decay widths
\begin{eqnarray}
\Gamma(h_{c}\to J/\psi \pi^+\pi^-)&=&\left\{42.6521(g_c-g_d)^2+3.0694(g_c-g_d)(g_f-g_h)+0.1978(g_f-g_h)^2 \right\} \,\rm{KeV}\, ,\nonumber\\
\Gamma(\chi_{c1}\to \eta_c \pi^+\pi^-)&=&\left\{340.7208(g_c+g_d)^2+40.3081(g_c+g_d)(g_f+g_h)+2.0529(g_f+g_h)^2 \right\} \,\rm{KeV}\, , \nonumber\\
\end{eqnarray}
depend heavily on the relative sign of the coupling constants. If $|g_c|\sim |g_d|$ and $|g_f|\sim |g_h|$, the special combinations $g_c \pm g_d$ and $g_f \pm g_h$ can lead to large augment or depression in those two-pion transitions.    In the case that the coupling constants $g_c$ and $g_d$ ($g_f$ and $g_h$) have  the same sign, the prediction is consistent with the value from the QCD multipole expansion \cite{Voloshin1209}.
We can fit the coupling constants $g_c$, $g_d$, $g_f$ and $g_h$ to the experimental data at the BESIII, KEK-B, RHIC, $\rm{\bar{P}ANDA}$ and LHCb
  in the future, and obtain quantitative predictions.

\section{Conclusion}
In this article, we study the two-pion decays of the $P$-wave to $S$-wave charmonium
transitions with the heavy meson  effective theory, and make qualitative  predictions
for ratios among the two-pion decay widths. We can confront the  decay widths with precise experimental data in the future to fit the coupling constants
and   obtain quantitative predictions.
\section*{Acknowledgement}
This  work is supported by National Natural Science Foundation of
China, Grant Number 11075053,   and the
Fundamental Research Funds for the Central Universities.

\end{document}